\RequirePackage{fix-cm}
\documentclass[twocolumn,epjc3]{svjour3}  
\smartqed  % flush right qed marks, e.g. at end of proof
\RequirePackage{graphicx}
\usepackage{epsfig,subfigure,latexsym}
\usepackage{amsmath,amssymb}
\usepackage{xcolor}
\usepackage{braket}

\usepackage{hyperref}
\journalname{Eur. Phys. J. C}

\newcommand {\bea}{\begin{eqnarray}}
\newcommand {\eea}{\end{eqnarray}}
\newcommand {\be}{\begin{equation}}
\newcommand {\ee}{\end{equation}}

\begin{document}
	\def\Journal#1#2#3#4{{\it #1} {\bf #2}, #3 (#4) }
\def\RPP{{Rep. Prog. Phys}}
\def\PRC{{Phys. Rev. C}}
\def\PRD{{Phys. Rev. D}}
\def\PRB{{Phys. Rev. B}}
\def\PRA{{Phys. Rev. A}}
\def\ZPA{{Z. Phys. A}}
\def\NPA{{Nucl. Phys. A}}
\def\NPB{{Nucl. Phys. B}}
\def\JPG{{J. Phys. G }}
\def\PRL{{Phys. Rev. Lett.}}
\def\PR{{Phys. Rep.}}
\def\PLB{{Phys. Lett. B}}
\def\AP{{Ann. Phys. (N.Y.)}}
\def\EPJA{{Eur. Phys. J. A}}
\def\EPJC{{Eur. Phys. J. C}}
\def\NP{{Nucl. Phys.}}  
\def\RMP{{Rev. Mod. Phys.}}
\def\IJMPE{{Int. J. Mod. Phys. E}}
\def\IJMPD{{Int. J. Mod. Phys. D}}
\def\AJ{{Astrophys. J.}}
\def\AJL{{Astrophys. J. Lett.}}
\def\AA{{Astron. Astrophys.}}
\def\ARAA{{Annu. Rev. Astron. Astrophys.}}
\def\MPLA{{Mod. Phys. Lett. A}}
\def\ARNPS{{Annu. Rev. Nucl. Part. Sci.}}
\def\LRR{{Living. Rev. Rel.}}
\def\CQG{{Classical Quantum Gravity}}
\def\RAS{{Mon. Not. R. Astron. Soc.}}
\def\JPA{{J.Phys. A}}
\def\ATMP{{Adv. Theor. Math. Phys}}
\def\GRG{{Gen. Relativ. Gravit}}

\title{ Minimal Length, Nuclear Matter, and Neutron Stars %\thanksref{t1}
}


\author{I. Prasetyo\thanksref{e1,addr2,addr3}
        \and
        I. H. Belfaqih\thanksref{e2,addr2,addr3}
        \and
        A. B. Wahidin\thanksref{e3,addr4,addr5}
        \and 
        A. Suroso\thanksref{e4,addr6}
        \and
        A. Sulaksono\thanksref{e5,addr1} %etc.
}

\thankstext{e1}{e-mail:ilham.prasetyo@sampoernauniversity.ac.id }
\thankstext{e2}{e-mail:idrus.belfaqih@sampoernauniversity.ac.id }
\thankstext{e3}{e-mail:alka.wahidin@sap.itera.ac.id }
\thankstext{e4}{e-mail:agussuroso@fi.itb.ac.id }
\thankstext{e5}{e-mail:anto.sulaksono@sci.ui.ac.id}


\institute{Department of General Education, Faculty of Art and Sciences, Sampoerna University, Jakarta 12780, Indonesia\label{addr2}
            \and
            IoT and Physics Lab, Sampoerna University, Jakarta 12780, Indonesia \label{addr3}
            \and
            Prodi Sains Atmosfer dan Keplanetan, Jurusan Sains Institut Teknologi Sumatera, Lampung Selatan 35365, Indonesia\label{addr4}
            \and
            Observatorium Astronomi ITERA Lampung, Institut Teknologi Sumatera, Lampung Selatan 35365, Indonesia\label{addr5}
            \and
            Theoretical High Energy Physics Research Division, Institut Teknologi Bandung, Jl. Ganesha 10 Bandung 40132, Indonesia\label{addr6}
			\and
			Departemen Fisika FMIPA Universitas Indonesia, Kampus UI, Depok 16424, Indonesia \label{addr1}
       }

\date{Received: date / Accepted: date}

\maketitle
 
\begin{abstract}
In this paper, we employ one variant of the Generalized Uncertainty Principle (GUP) model, i.e., the Kempf-Mangano-Mann (KMM) model, and discuss the impact of GUP on the EoS of nuclear and neutron star matter based on the Relativistic Mean Field (RMF) model. We input the result in the Serrano-Li\v{s}ka (SL) gravity theory to discuss the corresponding Neutron Star (NS) properties. We have shown that the upper bound for the GUP parameter from nuclear matter properties is $\beta \leq 2\times10^{-7}$ MeV$^{-2}$. If we used this $\beta$ upper bound to calculate NS matter, and considering  SL parameter $\tilde{c}$ as an independent parameter, we have found that the upper bound for the SL parameter, which modifies the Einstein field equation, is $\tilde{c} \leq 10^7$ m$^2$. This beta upper bound is determined by considering the anisotropy magnitude smaller than the pressure magnitude. By employing $\beta =2\times10^{-7}$ MeV$^{-2}$ and $\tilde{c} = 10^7$ m$^2$, we obtain the mass-radius relation that satisfies NICER data for both PSR J0740+6620 (whose mass is $\sim 2.1M_\odot$) and PSR J0030+0451 ($M\sim 1.4M_\odot$). Our GUP parameter upper bound perfectly matches the constraint from $^{87}$Rb cold-atom-recoil experiment. If we consider that the same strength from the additional logarithmic term in the entropy from both GUP and SL model are dependent, for $\beta < 2\times10^{-7}$ MeV$^{-2}$, it is clear that SL parameter lower bound is $\tilde{c} > -16\times 10^{-34}$ m$^2$. The magnitude of this bound is $10^{-40}$ smaller than the upper bound magnitude of SL parameter considering as independent parameter i.e., $\tilde{c} \leq 10^7$ m$^2$. 
%It means that the upper bound for the GUP parameter tends to increase as we go from the quantum regime to the gravity regime.
% \keywords{First keyword \and Second keyword \and More}
\end{abstract}

 \section{Introduction}
\label{sec_intro}

The grand theory of quantum gravity remains an elusive theory to be discovered. Several models have been proposed, for example, loop quantum gravity and string theory. These models have a common notion, i.e., on the existence of minimal measurable length $O(l_p)$, with $l_p$ is Planck length. Therefore, the spatial distance cannot be reduced to a point. In other words, there is a scale where space is fuzzy. In string theory, for instance, one cannot demand the spatial resolution to be smaller than the fundamental length of the string \cite{twns,Venez}. In loop quantum gravity, on the other hand, the space granularity is inherited by the $SU(2)$ structure of the quantum theory, which results in the area and volume eigenvalues described by the angular momentum $j$ representation \cite{RS,Ashtekar1,Ashtekar2}. The consequence of incorporating a minimal length in quantum mechanics (QM) formulation is modifying the Heisenberg uncertainty principle known as the generalized uncertainty principle (GUP). There are many GUP model proposals. However, the first attempts to construct a GUP model were made by Kempf-Mangano-Mann \cite{KMM}. This model is quite popular and commonly known as the KMM model in the literature.
The commutator between the position and momentum operator in the KMM model depends on the quadratic momentum. Many authors have investigated the effect of GUP on  QM and gravitation systems. For example, we can find the discussions of the effect of GUP in the harmonic oscillator in \cite{Achim} and hydrogen atom in \cite{Brau}. We can also find the discussion that the GUP yields black-hole remnants in \cite{Chen}. At the same time, the corresponding impact on the density of states and cosmological constant is discussed in  \cite{Minic}. While Hawking's temperature is in \cite{Scardigli}, and astrophysical objects are in \cite{Matthew,Idrus}.
Furthermore, some works to incorporate the formalism of GUP into quantum field systems have also been reported recently. In particular, the procedures have been made to the standard electromagnetic field to describe the Casimir effect \cite{PF85,Nouicer}. Inspired by GUP, recently in \cite{Bosso} has also implemented the modified Poisson relation of the minisuperspace and midisuperspace variables of the Wheeler de Witt system. To this end, we need to highlight that through GUP, the universality of quantum gravity corrections could be found in any quantum mechanical system \cite{Das1}.

It is known that the obstacle behind a direct test of quantum gravity is the energy scale related to the Planck length ($l_p$), $E_p \approx$ $10^{19}$ GeV. This energy scale is outside of the capability of current experimental technology. Therefore, determining the upper bound values of the GUP parameter from a theoretical framework, from gravitational observations, and experimental framework using QM systems becomes very crucial as an indirect test of the quantum gravity effect. Many attempts have been performed to constrain the GUP parameter values from theoretical and all possible observations and experimental frameworks, including those from QM systems. See Refs. \cite{Scardigli:2019pme,Okcu:2021oke,Giardino:2020myz} and the references therein for the review of the recent upper bound constraint of the parameter of GUP models. Note that other attempts to find possible methods or experimental setups to constrain the GUP parameter. For example, they have been discussed in Refs.~\cite{Idrus,Addconst1,Addconst2,Addconst3,Addconst4,Addconst5,Addconst6,Addconst7,Addconst8}. We need to highlight the results of review report in \cite{Scardigli:2019pme,Okcu:2021oke,Giardino:2020myz}. The different theoretical frameworks and explicit calculations yield a value for quadratic GUP parameter in the same order, i.e., $\beta \approx 1$. The best upper bound on $\beta$ value from gravitational origin probe, if we allow for violation of the Equivalence Principle, gives $\beta < 10^{21}$~\cite{Ghosh:2013qra} while if the Equivalence Principle is preserved, the upper bound of $\beta$ prediction becomes larger~\cite{Feng:2016tyt}. While the more restricted upper-bound prediction than the ones of gravitational probes comes from QM system probes (harmonic oscillators), i.e., $\beta < 10^{6}$. Therefore, it is obvious that the challenge and issue is the large difference in the upper bound of GUP parameters between the one predicted by theoretical frameworks and the ones from observation or experimental measurements.

This fact motivates us to investigate whether the symmetric nuclear matter (SNM), pure neutron matter (PNM), and neutron stars (NSs)  within the relativistic mean-field approach (RMF) could constrain the free parameter KMM GUP model. Note that the nuclear matters are dense many-body QM systems. They might be an appropriate area to test the GUP because we have a relatively certain and sufficient number of experimental data to constrain the equation of state (EoS) from low to relatively high-density regions of nuclear matters and NSs are the most compact horizonless objects observed in the universe. Furthermore, the nuclear matters and finite nuclei properties  predicted by the standard RMF model are compatible with experiment and observation data. For detailed discussions related to SNM and PNM predicted by standard  RMF models and their compatibility with experimental data and NS properties predictions, please see Ref. \cite{SB2012-a,SB2012-b,SB2012-c,SB2012-d,SB2012-e} and the references therein. In this work, we will also show that when we consider GUP correction in the EoS of matter, the first law of thermodynamics is preserved only if the pressure of matter is anisotropic.   Many authors have already explored and investigated the impacts of GUP on the many aspects of white dwarf's properties. Please see Refs. ~\cite{Idrus,Gregoris2022} and the references therein.
On the other hand, as far as we know that the impacts of GUP on NS properties were only discussed in Ref.~\cite{Abac2021}. The authors use the linear RMF (Walecka) model and general relativity (GR) to describe EoS and gravity, respectively. In this work, we use the standard RMF model to describe the relative realistic NS EoS, and we also consider the impact of GUP in the gravity sector. Therefore, we use Serrano-Li\v{s}ka~\cite{Alonso-Serrano:2020dcz} gravity theory to describe NS properties. The field equation of Serrano-Li\v{s}ka~\cite{Alonso-Serrano:2020dcz} gravity theory was derived using the thermodynamics of spacetime formalism and taking into account the additional logarithmic area term in Bekenstein entropy. The later modification is predicted by some quantum gravity approaches, including GUP phenomenology. Furthermore, recently the same authors \cite{Alonso-Serrano:2021} have also shown that the Hawking radiation in  Schwarzchild black hole agrees with the one obtained by other quantum gravity approaches, both heuristic and rigorous, including GUP phenomenology.   

We organized this paper as follows: In section \ref{gup}, we briefly discuss the GUP model used in this work, i.e., the KMM model. In section \ref{RMF}, we discuss the impact of GUP on the EoS of nuclear and neutron star matter based on the RMF model. In section \ref{nsm}, we discuss the NS properties based on Serrano-Li\v{s}ka gravity theory. In section \ref{RAD}, we discuss the results. Finally, the conclusions of this work are given in Section ~\ref{sec_conclu}.

\section{Generalized uncertainty principle}
\label{gup}

This section will briefly discuss the generalized uncertainty principle, which will be implemented in this work. To include the effects of minimal length in quantum mechanics, one must impose a deformation of the Heisenberg algebra into quadratic functions of momentum operator~\cite{KMM}. The general form of such deformation in $n$ spatial dimension that satisfies rotational and translational invariance is \cite{PF85}
\begin{equation}
	[\hat{x}_i,\hat{p}_j]=i\hbar\left[f\left(\hat{p}^2\right)\delta_{ij}+g\left(\hat{p}^2\right)\hat{p}_{i}\hat{p}_{j}\right],
	\label{eq:equation1}
\end{equation}
with $f\left(\hat{p}^2\right)$ and $g\left(\hat{p}^2\right)$ are some generic functions of total momentum squared. In this paper, we will work within the initial Kempf-Mangano-Mann model, 
in which the correction is coming from the leading quadratic term  \cite{KMM}
. The functions have the form of
\begin{equation}
	f\left(\hat{p}^2\right)=1+\beta \hat{p}^2,\quad g\left(\hat{p}^2\right)=0.
	\label{eq:equation2}
\end{equation}
This model commonly known as quadratic GUP, and the Jacobi relation resulting  non-commutative space coordinates relation. Other model in \cite{Das2} include linear term in addition to quadratic term to ensures the fundamental commutative relation $[x_i,x_j]=[p_i,p_j]=0$. Notice that the dimension of $\beta$ is inverse momentum square, so we define a dimensionless GUP parameter ${\alpha_Q}^2=(\hbar/l_P)^2\beta$, where $l_P$ is the Planck length.

GUP is a good tool for analyzing quantum gravity correction at a low energy limit. One of its predictions was reproducing a general result in quantum gravity, namely logarithmic correction to the Bekenstein-Hawking relation. As shown in \cite{Barun,Kaul,Medved,Meissner} that GUP generated the same logarithmic correction with $\beta$ playing the role as the expansion coefficient. Heuristically, when a black hole absorbs a particle with energy $E$ and size $\Delta x$, the area increases by
\begin{equation}
	\Delta A \geq \frac{8\pi l_{P}^{2}E\Delta x}{\hbar c}\geq \frac{8\pi l_{P}^{2}\Delta p\Delta x}{\hbar}
	\label{eq:equation3}
\end{equation}
with $l_p$ is the Planck length. [Please see Refs.~\cite{Barun,Medved} for explaination on how Eq.~\eqref{eq:equation3} obtained from Eqs. \eqref{eq:equation1} and \eqref{eq:equation2}.] The modification appears through $\Delta p \Delta x$ and accordingly, by setting the uncertainty of the position around the Schwarzschild radius $\Delta x \approx 2\pi R_S$ following Refs.~\cite{Barun,Medved}, the lower bound of the increasing area within the GUP model then
\begin{equation}
	\Delta A_{\text{min}}\approx 4\pi l_{P}^{2}\lambda\left[1+\frac{\hbar^2\beta}{4\pi A}\right],
\end{equation}
where $\lambda$ is proportionality constant to be determined later. Since the minimum increasing of entropy is by one bit corresponding to $\Delta S_{\text{min}}=\ln 2$ and hence
\begin{equation}
	\frac{dS}{dA}=\frac{\Delta S_{\text{min}}}{\Delta A_{\text{min}}}=\frac{\ln 2}{4\pi l_{P}^{2}\lambda\left[1+\frac{\hbar^2\beta}{4\pi A}\right]},
\end{equation}
where by direct integration, one obtain
\begin{equation}
	S\left(A,\beta\right)=\frac{k_{B}c^3}{4G\hbar}\left[A-\frac{\hbar^2\beta}{4\pi}\ln\left(\frac{A}{4l_{P}^{2}}\right)\right],
\end{equation}
where the proportional constant is obtained by demanding the relation is reduced to the standard relation of Hawking-Bekenstein.

It was also shown heuristically that the associated Hawking evaporation is halted at the Planckian scale due to the modification of the Hawking temperature \cite{Chen}. This modification of black hole thermodynamics inspired the work in \cite{Alonso-Serrano:2020dcz} to construct the modified GR due to the presence of minimal area and modified Hawking temperature by utilizing the maximal vacuum entanglement hypothesis (MVEH) of Jacobson \cite{Jacobson}.

We can no longer introduce a plane-wave state by introducing deformation in the canonical commutation relation since it provides a state with precise certainty of position. However, one can still define a state which saturates the uncertainty in position, known as the maximally localized state \cite{KMM}. By choosing the quadratic GUP, the corresponding maximally localized state has the form of
\begin{equation}\label{coh}
	\psi\left(t,\vec{x}\right)=\frac{1}{\sqrt{4\pi^3}}e^{-i\left(\omega_pt-\vec{P}\left(p\right).\vec{x}\right)},
\end{equation}
with $N$ is a normalization factor, and the functions $\vec{P}(p)$ has the following relation
\begin{equation}\label{mom}
	\vec{P}(p)=\frac{\vec{p}}{\sqrt{\beta}p}\tan^{-1}\left(\sqrt{\beta p}\right).
\end{equation}
The associated deformed completion relation reads
\begin{equation}
	\int \frac{d^3 p}{1+\beta p^2}\ket{p}\bra{p}=1.
\end{equation}
We observe that the deformation factor  above suppressed the high momentum region. This deformed measure leads to non-local inner product of maximally localized state as~\cite{Nouicer,Matsuo}
\begin{eqnarray}
	\braket{\psi_{\vec{x}}|\psi_{\vec{y}}}&=&\frac{1}{4\pi^3}\int \frac{1}{1+\beta p^2}e^{i\left(\vec{y}-\vec{x}\right).\vec{P}\left(p\right)}\nonumber \\
	&=&\frac{1}{8 \pi^3\beta}\frac{\sin\left(\frac{\left(x-y\right)\pi}{2\hbar \sqrt{\beta}}\right)}{\frac{x-y}{2\hbar \sqrt{\beta}}-\left(\frac{x-y}{2\hbar \sqrt{\beta}}\right)^2}\nonumber \\
	&=&\tilde{\delta}\left(\vec{x}-\vec{y}\right).
\end{eqnarray}
This expression yields in the modified Dirac delta functions that show non-locality. In the limit of $\beta\rightarrow0$ the standard Dirac delta will be recovered.

To describe nuclear matter, we have to provide how the GUP formalism is implemented on the fermionic matter. Since the momentum operator is modified here as $\hat{p_i}\rightarrow\hat{P_i}\left(p\right)$ as in Eq. (\ref{mom}), then the modified Dirac equations has the form of
\begin{equation}
	\left(\gamma^\mu \hat{P}_{\mu}-m\right)\psi_{\vec{x}}\left(t\right)=0,
\end{equation}
whereby using the maximally localized state (\ref{coh}), we obtain the modified dispersion relation as
\begin{eqnarray}
	\omega_{p}^{2}&=&P^2+m^2 \nonumber \\
	&=&\frac{1}{\beta}\left[\tan^{-1}\sqrt{\beta}p\right]^2+m^2.
\end{eqnarray}
From this analysis, by following the standard procedures of second quantization, we expand the field operator in terms of maximally localized state and also use the deformed measure as
\begin{eqnarray}
	\hat{\Psi}\left(\vec{x},t\right)&=&\sum_{\alpha}\int\frac{d^3p}{1+\beta p^2}[~\hat{b}\left(\vec{p},\alpha\right)\psi_{\vec{x}}\left(t\right)u\left(\vec{p},\alpha\right)\nonumber\\&+&\hat{d}^\dagger\left(\vec{p},\alpha\right)\psi_{\vec{x}}^{\dagger}\left(t\right)v\left(\vec{p},\alpha\right)~],
\end{eqnarray}
where we have chosen the normalization condition of the form
\begin{eqnarray}
	\sum_{\alpha=1}^{2}u^{\dagger}\left(\vec{p},\alpha\right)u\left(\vec{p},\alpha\right)&=&2, \nonumber \\
	\sum_{\alpha=1}^{2}\bar{u}^{\dagger}\left(\vec{p},\alpha\right)u\left(\vec{p},\alpha\right)&=&\frac{2m}{\omega_p},
\end{eqnarray}
and the anti-commutation of the annihilation and creation operator reads
\begin{eqnarray}
	\{\hat{b}\left(\vec{p},\alpha\right),\hat{b}^{\dagger}\left(\vec{q},\beta\right)\}&=&\{\hat{b}\left(\vec{p},\alpha\right),\hat{b}^{\dagger}\left(\vec{q},\beta\right)\}\nonumber\\&=&\left(2\pi\right)^3\delta_{\alpha\beta}\delta\left(\vec{p}-\vec{q}\right).
\end{eqnarray}
From the corresponding field operator expansion, normalization condition, and the anticommutation relation, we obtained the modified equal time  anticommutation field operator relation as
\begin{equation}
	\{\hat{\Psi}_a\left(\vec{x},t\right),\hat{\Psi}_b^{\dagger}\left(\vec{y},t\right)\}=\delta_{ab}\tilde{\delta}\left(\vec{x}-\vec{y}\right).
\end{equation}
So the relation is modified by replacing the standard Dirac delta with the deformed function. The above construction can be reduced to the standard results by taking the limit of $\beta\rightarrow0$. The following sections will implement this construction for nuclear matter and neutron star matter within the RMF models.

\section{GUP in nuclear and neutron star matters}
\label{RMF}

Nuclear matter and finite nuclei can be described by RMF models. The Lagrangian density of RMF models is defined as~\cite{SB2012-a,SB2012-b,SB2012-c,SB2012-d,SB2012-e}
\begin{equation}
	\label{eq:LagrangianTotal}
	\mathcal{L} = \mathcal{L}_N + \mathcal{L}_M  + \mathcal{L}_{int},
\end{equation}
which contain the contribution of free nucleons, mesons and interactions term. The free nucleons in finite nuclei has form
\begin{equation}
	\label{eq:LagrangianBarion}
	\mathcal{L}_N = \sum_{N=n,p}\overline{\psi}_N\left(i\gamma_{\mu}\partial^{\mu} - M_N\right)\psi_N.
\end{equation}
Here the sum is taken over all nucleons $N$ in nuclei. Nuclear matter is a thermodynamic limit of finite nuclei. Therefore, in this limit, $N \rightarrow \infty$ and volume goes to infinity, but the densities are finite. Therefore, in this limit, we replace $\sum_{N}$ with $\int d^3k$. Note that the interactions between nucleons are mediated by the exchange of scalar-isoscalar ${\sigma}$, vector-isoscalar  ${\omega}$, and vector-isovector ${\rho}$, mesons, respectively. Furthermore, the corresponding mesons have self-interactions. The interaction Lagrange density for finite nuclei  taken following form \cite{HP2001}:
\begin{eqnarray}
	\label{eq:LagrangianInteraksi}
	\mathcal{L}_{int} &=& \sum_{N=p,n}g_{\sigma}\sigma\overline{\psi}_N\psi_N - \sum_{N=p,n}g_{\omega}V_{\mu}\overline{\psi}_N\gamma^{\mu}\psi_N \nonumber \\
	&&-\sum_{N=p,n}g_{\rho}{\bold b}_{\mu}\cdot\overline{\psi}_{N}\gamma^{\mu} {\boldsymbol \tau}\psi_N - \frac{1}{3} b_2\sigma^3 - \frac{1}{4}b_3\sigma^4 \nonumber \\
	&& +\frac{1}{4}c_3\left(V_{\mu}V^{\mu}\right)^2 +\frac{1}{2}d_4\left(V_{\mu}V^{\mu}\right) \left({\mathbf b}^{\nu}\cdot {\mathbf b}_{\nu}\right)\nonumber \\
	&& + d_2\sigma \left(V_{\mu}V^{\mu}\right) +f_2\sigma \left({\mathbf b}^{\mu}\cdot {\mathbf b}_{\mu}\right)+ \frac{1}{2}d_3\sigma^2\left(V_{\mu}V^{\mu}\right).\nonumber\\
\end{eqnarray}
For free mesons, the Lagrangian density is as follows
\begin{equation}
	\label{eq:LagrangianMeson}
	\mathcal{L}_M = \mathcal{L}_{\sigma} + \mathcal{L}_{\omega} + \mathcal{L}_{\rho}, \\
\end{equation}
where the explicit form of each term is 
\begin{eqnarray}
	\mathcal{L}_{\sigma} &=& \frac{1}{2}\left(\partial_{\mu}\sigma\partial^{\mu}\sigma - m_{\sigma}^2\sigma\right) \label{eq:SigmaMeson}, \\
	\mathcal{L}_{\omega} &=& -\frac{1}{2}\left(\frac{1}{2}\omega_{\mu\nu}\omega^{\mu\nu} - m_{\omega}^2V_{\mu}V^{\mu}\right) \label{eq:OmegaMeson}, \\
	\mathcal{L}_{\rho} &=& -\frac{1}{2}\left(\frac{1}{2}{\boldsymbol \rho}_{\mu\nu}\cdot {\boldsymbol \rho}^{\mu\nu} - m_{\rho}^2{\mathbf b}_{\mu} \cdot {\mathbf b}^{\mu}\right) \label{eq:RhoMeson}.
\end{eqnarray}
Within the mean field approximation, $\sigma$, ${V}^{\mu}({V}_0, 0)$, and ${\mathbf b}^{\mu}({\mathbf b}_0, 0)$ are ${\sigma}$, ${\omega}$, and ${\rho}$ fields, respectively, and $\omega_{\mu\nu}$ and ${\boldsymbol \rho}_{\mu\nu}$ are the anti-symmetric tensor fields of ${\omega}$ and ${\rho}$ meson. Note that for the case NS matter, the $\beta$-stability condition should be satisfied. Therefore, the electrons and muons (leptons) should be exist in the NS matter. The contribution of non-interacting leptons to the total Lagrangian density is as follows
\begin{equation}
	\label{eq:LagrangianLepton}
	\mathcal{L}_L = \sum_{L=e,\mu}\overline{\psi}_L\left(i\gamma_{\mu}\partial_{\mu} - m_L\right)\psi_L.
\end{equation}

In the following, we will discuss the impact of the phase space deformation due to GUP on the nuclear matter and NS.  Using the RMF calculation procedure \cite{Glendenning}, we obtained the modified nucleon number densities for nuclear matter due to phase space deformation caused by GUP as
\bea
\label{NLnumberden}
\rho^*_N=\frac{2}{{(2 \pi)}^3} \int_0^{k_{fN} } \frac{d^3k}{\left(1+\beta k^2\right)^2},~~~N=p,n.
\eea  
Similarly, scalar number densities for protons and neutrons are expressed as follows
\begin{equation}
	\label{NLscalar_numberden}
	\rho^*_{s~N}=\frac{2}{{(2 \pi)}^3}\int_0^{k_{fN}}\frac{M^*_N}{\omega_N}\frac{d^3k}{\left(1+\beta  k^2\right)^2},
\end{equation}  
where \(\omega_N = \sqrt{\frac{1}{\beta}{\left(\tan^{-1} [\sqrt{\beta} k]\right)}^2+M_N^{*~2}}\) and $M^*_N=M_N+g_\sigma \sigma$.
The nucleon contributions in energy density are as follows
\begin{eqnarray}
	\label{nuceden}
	\epsilon_N^*= \frac{2}{ {(2 \pi)}^3} \int_0^{k_{fN} } \omega_N \frac{d^3k}{\left(1+\beta  k^2\right)^2}.
\end{eqnarray}
The explicit expressions for $P_{rN}^*$ is
\begin{eqnarray}
	\label{nucpress}
	P^*_{rN}= \frac{2}{{(2 \pi)}^3}\int_0^{k_{fN}}\frac{k^2}{\omega_N}\frac{d^3k}{\left(1+\beta  k^2\right)^2}.
\end{eqnarray}

Due to the physical fact that $\beta$ should be a small number, we can expand the number density in  Eq.~(\ref{NLnumberden}), the scalar density in Eq.~(\ref{NLscalar_numberden}), the energy density in Eq.~(\ref{nuceden}) and the radial pressure in Eq.~(\ref{nucpress}) in respect to $\beta$. If we take only up to the first order of $\beta$, we can obtain simple analytical expressions. The  $\rho_{N}^*$ can be approximated as
\begin{eqnarray}
	\rho_{N}^* = \rho_{N} + \beta ~\Delta\rho_{N},
\end{eqnarray}
where
\begin{eqnarray}
	\rho_{N} &=&  \frac{1}{3\pi^2} k_{fN}^3,\nonumber\\
	\Delta\rho_{N} &\approx&-\frac{2}{5\pi^2} k_{fN}^5,
\end{eqnarray}
while $\rho_{s~N}^*$ now becomes
\begin{eqnarray}
	\rho_{s~N}^* = \rho_{s~N} + \beta~\Delta\rho_{s~N},
\end{eqnarray}
with 
\begin{eqnarray}
	\rho_{s~N}&=&\frac{M_N^*}{2 \pi^2}\Bigg\{ k_{fN}\sqrt{k_{fN}^{2}+M_N^{*^2}}\nonumber\\
	&&-M_N^{*^2}~\ln\Bigg[\frac{k_{fN}+\sqrt{k_{fN}^{2}+M_N^{*^2}}}{M_N^*}\Bigg]\Bigg\},
\end{eqnarray}
and the GUP correction on $\rho_{s~N}^*$ can be approximated as
\begin{eqnarray}                      
	\Delta\rho_{s~N} &\approx& \frac{M_N^*}{24 \pi^2}\Bigg\{ \frac{\Big[-10 k_{fN}^{5}+k_{fN}^{3}M_N^{*^2}    +3k_{fN} M_N^{*^4}\Big]}{\sqrt{k_{fN}^{2}+M_N^{*^2}}}\nonumber\\
	&&-3M_N^{*^4}~\ln\Bigg[\frac{k_{fN}+\sqrt{k_{fN}^{2}+M_N^{*^2}}}{M_N^*}\Bigg]\Bigg\}.
\end{eqnarray}
Similarly for 	$\epsilon_N^*$, we have
\begin{equation}
	\epsilon_N^* = \epsilon_N + \beta \Delta \epsilon_N ,
\end{equation}
with 
\begin{eqnarray}
	\label{eq:Energy Density nucleon} 
	\epsilon_N &=&  \frac{1}{8\pi^2} \biggl\{ k_{fN} \sqrt{k_{fN}^2 + M_N^{*^2}} \biggl[2k_{fN}^2 + {M_N^*}^2 \biggr] \nonumber \\
	&& -{M_N^*}^4 \ln \biggl[ \frac{k_{fN} + \sqrt{k_{fN}^2 + {M_N^*}^2}}{M_N^*} \biggr] \biggr\},
\end{eqnarray}
and the GUP correction on the GUP correction on $\epsilon_{N}^*$ is 
\begin{eqnarray} 
	\Delta \epsilon_N &\approx&  +\frac{1}{144\pi^2} \biggl\{ k_{fN} \sqrt{k_{fN}^2 + M_N^{*^2}} \nonumber\\&\times& \biggl[-56 k_{fN}^4-2k_{fN}^2 {M_N^*}^2+3 {M_N^*}^4 \biggr] \nonumber \\
	&& - 3 {M_N^*}^6 \ln \biggl[ \frac{k_{fN} + \sqrt{k_{fN}^2 + {M_N^*}^2}}{M_N^*} \biggr] \biggr\}.
\end{eqnarray}
For the radial pressure of nucleon $P_{r N}^{*}$, we have 
\begin{equation}
	\label{eq:Pold}
	P_{r N}^{*} = P_{rN} + \beta \Delta P_{r N} ,
\end{equation}
with
\begin{eqnarray}
	\label{eq:Pressure isotropic}
	P_{r N} &=&  \frac{1}{8\pi^2} \biggl\{ k_{fN} \sqrt{{k_{fN}}^2 + {M_N^*}^2} \biggl(2{k_{fN}}^2 - 			3{M_N^*}^2 	\biggr) \nonumber \\
	&&+ 3{M_N^*}^4 \ln \biggl[ \frac{k_{fN} + \sqrt{{k_{fN}}^2 + {M_N^*}^2}}{M_N^*} \biggr] \biggr\},
\end{eqnarray}
and the corresponding GUP correction in $P_{r N}^{*}$ is as follows
\begin{eqnarray}
	\label{eq:Pressure Radial Proton}
	\Delta P_{rN} &\approx&      +\frac{1}{144\pi^2} \left\{- 15 {M_N^*}^6 \ln \biggl[ \frac{k_{fN} + \sqrt{k_{fN}^2 + {M_N^*}^2}}{M_N^*} \biggr]  \right.\nonumber\\&+&\left. Z(N,M^*_N) \right\}, 
\end{eqnarray}
with
\begin{eqnarray}
Z(N,M^*_N)&=&	 \left[   -40 k_{fN}^7-2k_{fN}^5 {M_N^*}^2     + 5k_{fN}^3 {M_N^*}^4   \right.\nonumber\\&+& \left.15k_{fN} {M_N^*}^6 \right] \left\{k_{fN}^2 + M_N^{*^2}\right\}^{-1/2}.
\end{eqnarray}
Note, for leptons, the expressions are similar to those of nucleons. However, in lepton cases $\rho_{L}^*= \rho_{N}^* (N\rightarrow L,M_N^* \rightarrow m_L)$,  $\epsilon_{L}^*= \epsilon_{N}^* (N\rightarrow L,M_N^* \rightarrow m_L)$, and $P_{rL}^*= P_{rN}^* (N\rightarrow L,M_N^* \rightarrow m_L)$, with L=e, $\mu$. 

In this way, we have the total number density $\rho$  in NS matter as 
\begin{equation}
	\rho = \sum_{N=n,p} \rho^*_N + \sum_{L=e,\mu} \rho^*_L,
	\label{rhotot}
\end{equation}
and scalar density $\rho_s$ is  
\begin{equation}
	\rho_s = \sum_{N=n,p} \rho^*_{sN} + \sum_{L=e,\mu} \rho^*_{sL},
\end{equation}
The total energy density $\epsilon$ can be expressed as follows
\begin{equation}
	\label{eden}
	\epsilon=\sum_{N=n,p}\epsilon_N^*+g_\omega (\rho^*_p+\rho^*_n)+\frac{1}{2}g_\rho (\rho^*_p-\rho^*_n)+U+\sum_{L=e,\mu}\epsilon_L,
\end{equation}
where the meson contribution still takes following form
\begin{eqnarray}
	U&=&\frac{1}{2}m_s^2 \sigma^2-\frac{1}{2}m_\omega^2 V_0^2   -\frac{1}{2}m_\rho^2 b_0^2
	\nonumber\\ && + \frac{1}{3}b_2\sigma^3+\frac{1}{4}b_3 \sigma^4-\frac{1}{4}c_1 V_0^4-\frac{1}{2} d_4 b_0^2 V_0^2\nonumber \\ && - d_2\sigma V_0^2-f_2\sigma b_0^2- \frac{1}{2}d_3\sigma^2V_0^2.
\end{eqnarray}
The total radial pressure 	$P_{r}$ is 
\begin{equation}
	\label{press}
	P_{r}=\sum_{M=n,p}\frac{1}{3}P^{*}_{rN}-U+\sum_{L=e,\mu}\frac{1}{3}P_{rL}.
\end{equation}
Based on these pressures and energy densities, and assuming we know the Fermi momentum of each constituent, we can calculate the EoSs of each corresponding type of matter. Note that, for PNM, we only required $k_{fn}$ to determine the EoS, whereas for SNM, we need $k_{fn}=k_{fp}$ as input to calculate the EoS. In addition, we need to note that the most crucial SNM parameter is the binding energy at the saturation density of SNM ($E/N$). Other nuclear-matter isoscalar properties at saturation density can be derived from the binding energy $E (\rho)$ using following expressions
\bea
K_0 &=& 9 \rho_0^2 \frac{d^2 E/N(\rho)}{d\rho^2} |_{\rho=\rho_0},
\label{Eq:NMISCR}
\eea 
while in the isovector sector of nuclear matter, the symmetry energy at the saturation density $J$ plays a role similar to that of the binding energy isoscalar sector. Other nuclear-matter isovector properties at saturation density of SNM can be derived from $J (\rho)$ and are given by the following relations
\bea
L &=& 3 \rho_0 \frac{d J (\rho)}{d\rho} |_{\rho=\rho_0},\nonumber\\
K_{\rm sym} &=& 9 \rho_0^2 \frac{d^2 J (\rho)}{d\rho^2} |_{\rho=\rho_0},\nonumber\\
\label{Eq:NMIVEC}
\eea 
For the EoS of the NS core, we need to apply $\beta$ stability and the neutrality conditions to determine the Fermi momentum of each constituent. In the following, we will discuss how to obtain the Fermi momentum of each constitient in NS. The chemical potential of each constitient $\mu_i^*$= $\frac{d \epsilon}{d\rho^*_i}$ with $i$= p, n, e, and $\mu$, can be approximated as
$\mu_i^*$ $\approx$ $\mu_i + \beta \Delta \mu_i$.  The explicit expressions are 
\begin{align}
	\mu_n^* &\approx \sqrt{k_{fn}^2 + M_n^{*^2}}-\beta      \frac{1}{3} \frac{k_{fn}^4}{\sqrt{k_{fn}^2 + M_n^{*^2}}} +g_{\omega} V_0-\frac{1}{2} g_{\rho} b_0, \nonumber\\ 
	\mu_p^* &\approx \sqrt{k_{fp}^2 + M_p^{*^2}}-\beta       \frac{1}{3} \frac{k_{fp}^4}{\sqrt{k_{fp}^2 + M_p^{*^2}}} +g_{\omega} V_0+\frac{1}{2} g_{\rho} b_0, \nonumber\\
	\mu_e^* &\approx \sqrt{k_{fe}^2 + m_e^{  2}}-\beta       \frac{1}{3} \frac{k_{fe}^4}{\sqrt{k_{fe}^2 + m_e^{  2}}}, \nonumber\\
	\mu_\mu^* &\approx \sqrt{k_{f\mu}^2 + m_{\mu}^{  2}}-\beta        \frac{1}{3} \frac{k_{f\mu}^4}{\sqrt{k_{f\mu}^2 + m_{\mu}^{  2}}}. 
\end{align}
To this end, by substituting all of the densities and chemical potentials to the conditions of $\beta$ stability and neutrality i.e., 
\begin{eqnarray}
	\rho^*_p&=&\rho^*_e+\rho^*_{\mu},\nonumber\\
	\mu^*_{\mu}&=&\mu^*_n+\mu^*_{p},\nonumber\\
	\mu^*_{\mu}&=&\mu^*_e,\nonumber\\
	\rho_B&\equiv&\rho_p^*+\rho_n^*,
\end{eqnarray}
and the approximate form of Fermi momentum is 
\begin{equation}
	k_{fi} \approx {(3 \pi^2 \rho_i^* )}^{(1/3)} + \beta (\frac{6}{5} \pi^2 \rho_i^* ),
	\label{kfapprox}
\end{equation}
we obtain the electron fraction $Y_e=\frac{\rho_e^*}{\rho_B}$ and proton fraction as  $Y_p=\frac{\rho_p^*}{\rho_B}$ as
\begin{eqnarray}
	Y_e &=& \frac{1}{3\pi^2 \rho_B}  \left[ {(\mu_e^{*^2} -m_e^{2})}{\left(1+\beta\tfrac{ 8}{15}(\mu_e^{*^2} -m_e^{2})\right)}   \right]^{3/2},  \nonumber\\ 
	\label{ye}
\end{eqnarray}
and
\begin{eqnarray}
	Y_p &=& \frac{1}{3\pi^2 \rho_B}  \left[ {(\mu_e^{*^2} -m_e^{2})}{\left(1+\beta\tfrac{ 8}{15}(\mu_e^{*^2} -m_e^{2})\right)}   \right]^{3/2} \nonumber\\ 
	&& +\frac{1}{3\pi^2 \rho_B}\left[ {(\mu_e^{*^2} -m_\mu^{2})}{\left(1+\beta\tfrac{8 }{15}(\mu_e^{*^2} -m_\mu^{2})\right)}  \right]^{3/2},\nonumber\\ 
	\label{yp}
\end{eqnarray}
while the explicit expression of $\mu_e^{*^2}$ is
\begin{eqnarray}
	\mu_e^* &=& X_n - X_p +\frac{\tfrac{1}{2}g_\rho^2[1-2Y_p]\rho_B^{2/3}}{m_\rho^2+2f_2\sigma+g_3\sigma^2+g_4\omega_0^2} +\beta \frac{4}{15}\rho_B^{1/3}
	\nonumber\\ &\times&\left[ \frac{(3\pi^2[1-Y_p])^{4/3} \rho_B^{2/3}}{X_n} 
	%\right.\nonumber\\&&\left.
	- \frac{(3\pi^2 Y_p)^{4/3} \rho_B^{2/3}}{X_p} \right] ,\nonumber\\
	\label{mue}
\end{eqnarray}
with $X_n=\sqrt{(3\pi^2[1-Y_p])^{2/3}+\frac{M_n^{*2}}{\rho_B^{2/3}}}$ and\\ $X_p = \sqrt{(3\pi^2 Y_p)^{2/3}+\frac{M_p^{*2}}{\rho_B^{2/3}}}$.
Eqs. (\ref{ye}-\ref{mue}) are solved self-consistently to obtain $Y_e$ and $Y_p$. Then we also have $Y_n=1-Y_p$, and by using Eq. (\ref{kfapprox}) and $\mu^*_{\mu}=\mu^*_e$ condition, we can obtain Fermi momentum of each particle in NS matter. We need to note that due to Eq. (\ref{rhotot}), the total chemical potential is
\begin{eqnarray}
	\mu (\beta) &=& \sum_{N=n,p} \mu^*_N + \sum_{L=e,\mu} \mu^*_L,\nonumber\\
	&=&\mu (\beta=0) +\beta \Delta \mu.
\end{eqnarray}
According to first law of thermodynamic, the total chemical potensial should be satisfied the following relation 
\begin{equation}
	\mu = \frac{\epsilon+P_r}{\rho}.
	\label{mutot}
\end{equation}
It is well known for standard RMF model where here, it is equal to $\beta$=0 case, Eq. (\ref{mutot}) is satisfied. However, we have found in the case $\beta\ne$0, this relation is not satisfied fully due to the fact that 
\begin{eqnarray}
	&&\mu (\beta)\rho(\beta) - \epsilon (\beta)   - P_r (\beta) \nonumber\\&\approx&   - \beta \left[ \rho (\beta=0)\Delta \mu +\mu(\beta=0) \Delta \rho-\Delta \epsilon -\Delta P_r \right]\nonumber\\&\neq& 0.
	\label{TD1isotropic}
\end{eqnarray}
We can resolve this issue by assuming that the pressure of NS  matter is slightly anisotropic due to GUP; namely, the radial pressure is not equal to tangential pressure. In general, the average pressure can be defined as $P=\frac{1}{3}[P_r+2P_t]$, where   $P_t$ is tangential pressure. If $\sigma\equiv P_r-P_t$, then  $P=P_r-\frac{2}{3} \sigma$. In order to satisfy the first law of thermodynamics, we should have 
\begin{equation}
	\mu (\beta)\rho(\beta) - \epsilon (\beta)    - P (\beta) =0
	\label{TD1unisotropic}
\end{equation}
Note for isotropic case $P=P_r$ the relation back to Eq. (\ref{TD1isotropic}) and Eq. (\ref{TD1unisotropic})  can be written as 
\begin{equation}
	\mu (\beta)\rho(\beta) - \epsilon (\beta)   -P_r (\beta)+\frac{2}{3}\sigma =0.
	\label{TD1unisotropic2}
\end{equation}
It is obvious the impact of anisotropic is encoded in $\sigma$. Therefore, if we choose 
\begin{equation}
	\sigma\equiv       -\frac{3}{2}\beta \left[ \rho (\beta=0)\Delta \mu  +\mu(\beta=0) \Delta \rho-\Delta \epsilon -\Delta P_r \right],
\end{equation}
the fundamental  first law of thermodynamics is satisfied, even for the case of nonzero $\beta$ value. 
We have found that by introducing GUP makes the relation between chemical potential as a function of energy density, pressure, and number density due to the consequence of the universal second law of thermodynamics is no longer fulfilled. This fact indicates that the corresponding EoS is no longer obey the second law of thermodynamics. One possible way that we know to restore the fulfilment the universal thermodynamics law is only by introducing ad hoc anisotropic pressure in matter. However, it seems that the solution of this issue needs more detailed investigation that could be beyond the scope of present work. Therefore, we will address this issue for future work.
Note that the explicit expressions of each contribution in $\sigma$ are
\begin{align}
	\rho &(\beta=0) = \sum_{N=p,n} \frac{k_{fN}^3}{3\pi^2}+\sum_{L=e,\mu} \frac{k_{fL}^3}{3\pi^2},\nonumber\\
	\mu&(\beta=0) = \sum_{N=p,n}\sqrt{k_{fN}^2 + M_N^{*^2}}+ 2 g_{\omega} V_0\nonumber\\&+\sum_{L=e,\mu}\sqrt{k_{fL}^2 + m_L^{2}},\nonumber\\
	\Delta \mu  &=   -\frac{1}{3}  \sum_{N=p,n}\frac{k_{fN}^4}{\sqrt{k_{fN}^2 + M_N^{*^2}}}  -\frac{1}{3}  \sum_{L=p,n}\frac{k_{fL}^4}{\sqrt{k_{fL}^2 + m_L^{2}}}\nonumber\\
	\Delta \rho  &= -\sum_{N=p,n} \frac{2 k_{fN}^5}{5 \pi^2} -\sum_{L=e,\mu} \frac{2k_{fL}^5}{5 \pi^2},
\end{align}
and
\begin{eqnarray}
	\Delta \epsilon  &=&      
	+\frac{1}{144\pi^2} \sum_{N=p,n}\biggl\{ k_{fN} \sqrt{k_{fN}^2 + M_N^{*^2}} \nonumber\\&\times& \biggl[-56 k_{fN}^4-2k_{fN}^2 {M_N^*}^2+3 {M_N^*}^4 \biggr] \nonumber \\
	&&- 3 {M_N^*}^6 \ln \biggl[ \frac{k_{fN} + \sqrt{k_{fN}^2 + {M_N^*}^2}}{M_N^*} \biggr] \biggr\} \nonumber\\
	&& + \frac{1}{144\pi^2} \sum_{L=e,\mu}\biggl\{ k_{fL} \sqrt{k_{fL}^2 + m_L^{2}}\nonumber\\&\times&  \biggl[-56 k_{fL}^4-2k_{fL}^2 {m_L}^2+3 {m_L}^4 \biggr] \nonumber \\
	&& - 3 {m_L}^6 \ln \biggl[ \frac{k_{fL} + \sqrt{k_{fL}^2 + {m_L}^2}}{m_L} \biggr] \biggr\}
	\nonumber \\ &&+g_\omega (\Delta\rho_p+\Delta\rho_n)+\frac{1}{2}g_\rho (\Delta\rho_p-\Delta\rho_n)
	,
\end{eqnarray}
with $\Delta\rho_{i} =-(2/5\pi^2) k_{fi}^5 (i=N,L)$, while
\begin{eqnarray}
	\Delta P_r  &=&  +\frac{1}{144\pi^2} \sum_{L=e,\mu}\biggl\{ Z(L,m_L) \nonumber\\ &+&  - 15 {m_L}^6 \ln \biggl[ \frac{k_{fL} + \sqrt{k_{fL}^2 + {m_L}^2}}{m_L} \biggr]  \biggr\}\nonumber\\
	&&+\frac{1}{144\pi^2} \sum_{N=p,n}\biggl\{ Z(N,M^*_N) \nonumber\\ &+& - 15 {M_N^*}^6 \ln \biggl[ \frac{k_{fN} + \sqrt{k_{fN}^2 + {M_N^*}^2}}{M_N^*} \biggr]
	\biggr\}.
\end{eqnarray}
We will used $P_r$, $\epsilon$ and $\sigma$ of EoS of NS to calculate the properties of NS. Note that there are many RMF parameter sets proposed to explain the finite nuclei and nuclear matter properties. See Refs. ~\cite{SB2012-a,SB2012-b,SB2012-c,SB2012-d,SB2012-e} for details. However, here we use the BSP parameter set because the predictions of this parameter set are relatively compatible with the experimental data of finite nuclei and nuclear matter.  

\section{Neutron stars within Serrano-Li\v{s}ka gravity theory }
\label{nsm}

The field equation of Serrano-Li\v{s}ka~\cite{Alonso-Serrano:2020dcz} gravity theory  takes a modified unimodular gravity structure as follow
\begin{equation}
	S_{ab}-\tilde{c} S_{ac} S^c_b +\frac{\tilde{c}}{4} S_{cd} S^{cd} g_{ab} = \kappa_N t_{ab}, \label{eq:SL}
\end{equation}
where \( S_{ab} = R_{ab}-(1/4)g_{ab} R \), \( t_{ab} = T_{ab}-(1/4)g_{ab} T \), \(R=R_{ab}g^{ab}\), \(T=T_{ab}g^{ab}\), and \( \kappa_N = 8\pi G\). Note that the natural units \( c=\hbar=1 \) are used. The third term with constant \(\tilde{c} \equiv D l_p^2\) where $D$ a dimensionless constant, is related to the logartithmic term in entropy correction.  Note that $\tilde{c}$ has a physical units of m$^2$. This equation has another constraint such that the energy-momentum tensor $T_{ab}$ conservation is satisfied, i.e.
\begin{eqnarray}
	{1\over 4} R_{,b} + {\kappa_N \over 4} T_{,b} &=& \tilde{c} \left( S^{ac}_{~~;a} S_{bc} + S^{ac} S_{bc;a} \right) \nonumber\\& -& {\tilde{c} \over 4} \left( S_{cd} S^{cd} \right)_{,b}, \label{eq:conservation}
\end{eqnarray}
where the semi-colon symbol denotes covariant derivative and the comma symbol denotes partial derivative.

Because we want to study spherically symmetric NS, the Schwarzchild metric is used i.e., 
\begin{equation}
	ds^2 = -e^{2\alpha(r)} dt^2 + e^{2\nu(r)} dr^2 + r^2 d\Omega^2,
\end{equation}
where \(d\Omega\) the infinitesimal element of a 2-sphere. To describe the NS interior, the ideal fluid anisotropic energy-momentum tensor as follow is used
\begin{eqnarray}
	T^{ab} &=& \rho u^a u^b + P r^a r^b + (P-\sigma) (g^{ab} + u^a u^b - r^a r^b),\nonumber\\
\end{eqnarray}
where we denote that \( u^a = \delta^a_0 (-g_{00})^{-1/2} \) and \( r^a = \delta^a_r (g_{rr})^{-1/2} \). Therefore, we can obtain the non-zero components of \(t_{ab}\) as
\begin{eqnarray}
	t_{00} &=& -\frac{1}{4} (3\rho + 3P - 2\sigma) g_{00},\\
	t_{rr} &=& \frac{1}{4} (\rho + P + 2\sigma) g_{rr},\\
	t_{\theta\theta} &=& \frac{1}{4} (\rho + P - 2\sigma) g_{\theta\theta}, \\ 
	t_{\varphi\varphi} &=& t_{\theta\theta} \frac{g_{\varphi\varphi}}{g_{\theta\theta}}.
\end{eqnarray}

Furthermore, by asumming that $\tilde{c}$ is small, we can also obtain the aprroximate forms of $g_{ab}$, $R$, and $T$ in first order of Taylor expansion in $\tilde{c}$ as
\begin{eqnarray}
	g_{ab} &=& g^{(0)}_{ab} +\tilde{c} g^{(1)}_{ab},\\
	R &=& R^{(0)} +\tilde{c} R^{(1)}, \\
	T &=& T^{(0)} +\tilde{c} T^{(1)}.
\end{eqnarray}
Hence \(S_{ab} = S^{(0)}_{ab} + \tilde{c} S^{(1)}_{ab}\) and \(t_{ab} = t^{(0)}_{ab} + \tilde{c} t^{(1)}_{ab}\). Substituting these expressions into Eq. (\ref{eq:conservation}), we obtain
\begin{equation}
	R^{(0)} + \kappa_N T^{(0)} = 0.
\end{equation}
and
\begin{eqnarray}
	{1\over 4} R^{(1)}_{,b} + {\kappa_N \over 4} T^{(1)}_{,b} &=&  \left( S^{(0)~ac}_{~~~~;a} S^{(0)}_{bc} + S^{(0)~ac} S^{(0)}_{bc;a} \right)\nonumber\\ &-&{1\over 4} \left( S^{(0)}_{cd} S^{(0)~cd} \right)_{,b}  , \label{eq:constraint02}
\end{eqnarray}
for the \(\mathcal{O}(1)\) and \(\mathcal{O}(\tilde{c})\) terms, respectively.

Therefore by using the explicit expression of metric components, we can have the following non-zero components of the Ricci tensor as
\begin{eqnarray}
	R_{00} &=&  e^{2(\alpha-\nu)} \left[ \alpha'' + \alpha'^2 - \alpha'\nu' +{2\over r} \alpha' \right], \\
	R_{rr} &=& - \left[ \alpha'' + \alpha'^2 - \alpha'\nu' -{2\over r} \nu' \right], \\
	R_{\theta\theta} &=& e^{-2\nu} \left[ r(\nu'-\alpha') - 1 \right] + 1,\\
	R_{\varphi\varphi} &=& R_{\theta\theta} \sin^2\theta,
\end{eqnarray}
where the prime symbol denotes differentiation with respect to \(r\). These first order approximations make \(S_{ab}=0\) if \(a\neq b\). Now, since the \(\mathcal{O}(1)\) terms in Eq.~(\ref{eq:SL}) is just the usual unimodular gravity equation \( S^{(0)}_{ab} = \kappa_N t^{(0)}_{ab} \), then we substitute this into Eq.~(\ref{eq:constraint02}), which becomes 
\begin{eqnarray}
	\left( R^{(1)} + \kappa_N T^{(1)} \right)_{,b} &=& -\kappa_N^2 g^{(0)bb} T^{(0)}_{,b} t^{(0)}_{bb} \nonumber\\&+& 4\kappa_N^2 \sum_a t^{(0)aa} t^{(0)}_{ba;a} \nonumber\\
	&-&\kappa_N^2 \sum_{c,d} \left( t^{(0)}_{cd} t^{(0)cd} \right)_{,b}. \label{eq:constraint2}
\end{eqnarray}
Note that here for the tensor term with repeated (dummy) index,  we do not sum to all components like the usual Einstein summation convention. Using \(b=r\) and employing the fact that the NS EoS \(\rho\) satisfies \(\rho=\rho(P)\) relation, we obtain
\begin{equation}
	R^{(1)} + \kappa_N T^{(1)} = {\kappa_N}^2 \Lambda(r),
\end{equation}
with $\Lambda(r)$ satisfies an equation of motion from Eq. \eqref{eq:constraint2} namely
\begin{eqnarray}
	\Lambda' &\equiv& P'(\rho+P-2\sigma) \left(1 - {d\rho \over dP}\right) \nonumber\\
	&&+ {8\sigma\over r} (\rho+P-\sigma) + 2\sigma' (\rho+P). 
	\label{eq:Lambdaprime}
\end{eqnarray}
Here prime denotes differentiation with respect to $r$.
Then we can obtain the \(r\) component from the Bianchi identity namely
\begin{equation}
	P'=-\alpha'(\rho+P)-{2\sigma \over r}.
	\label{eq:pressureprime}
\end{equation}
The components of Eq.~(\ref{eq:SL}) can then be also rewritten as
\begin{eqnarray}
	G_{aa}-\kappa_N T_{aa} &+& {\tilde{c} \kappa_N^2\over 4} g_{aa} \Lambda(r) =  \tilde{c} \kappa_N^2 (t_{aa})^2 g^{aa}\nonumber\\&-& {\tilde{c} \kappa_N^2\over 4} g_{aa} \sum_{c} (g_{cc} t^{cc})^2 ,
\end{eqnarray}
Here, \(G_{ab}=R_{ab}-(1/2)g_{ab}R\). Setting \(a=0\) and by assuming \(e^{2\beta}= \left(1 - {2Gm(r)/r} \right)^{-1},\)
we can obtain
\begin{eqnarray}
	m' &=& 4\pi r^2 \tilde{\rho},
	\label{eq:massprime}
\end{eqnarray}
with
\begin{equation}
	\tilde{\rho} = 
	\rho + {\kappa_N \tilde{c}\over 4}
	\left\{
	\Lambda(r) - {3\over 2} (\rho+P)^2 +2\sigma(\rho+P)
	\right\}.
\end{equation}
Note that the correction appears in the second term. Setting \(a=r\), we also obtain
\begin{eqnarray}
	\alpha' &=& {G m\over r^2} 
	\left[1 -{4\pi r^3\over m}
	\tilde{P}   \right] 
	\left(1 - {2Gm(r)\over r} \right)^{-1},
	\label{eq:alphaprime}
\end{eqnarray}
with
\begin{equation}
	\tilde{P} = P - {\kappa_N \tilde{c}\over 4}
	\left(\Lambda(r) + {1\over 2} (\rho+P)^2 -2\sigma(\rho+P).
	\right)
\end{equation}

We numerically integrate Eqs.~(\ref{eq:Lambdaprime})-(\ref{eq:pressureprime}), and Eqs.~(\ref{eq:massprime})-(\ref{eq:alphaprime}), from the center of NS \(r=r_c\sim0\) to its surface \(r=R\). The boundary conditions at the center are \( P(r_c)=P_c, \rho_c = \rho(P_c), \sigma(r_c)=0, m(r_c)=m_c=(4/3)\pi \rho_c r_c^3, \) and \(\Lambda(r_c)=\Lambda_c\), with $\Lambda_c$ arbitrary.

The boundary conditions at the surface are \( P(r)=0, m(R)=M, \alpha(R)=(1/2)\ln(1-2GM/R),\) and \(\Lambda(0)=0\). The pressure at the center \(P_c\) is chosen arbitrarily. Both \(\alpha(r_c)\) and \(\Lambda(r_c)\) can be determined from the shooting method, i.e., by taking an arbitrary value of $\alpha(r_c)=\alpha_{c,old}$ and $\Lambda(r_c)=\Lambda_c$ at the center and we iterate the calculation with following relations
\begin{eqnarray}
	\alpha_{c,new}&=&\alpha_{c,old}-\left[
	\alpha(R)-(1/2)\ln(1-2GM/R)
	\right],\\
	\Lambda_{c,new}&=&\Lambda_{c,old}-\Lambda(R),
\end{eqnarray}
and follow by recalculating the equations again from the center. The looping stops when \(|\alpha_{c,new}-\alpha_{c,old}|<\varepsilon_\alpha\) and \(|\Lambda_{c,new}-\Lambda_{c,old}|<\varepsilon_\Lambda,\) with $\varepsilon_\alpha>0$ and $\varepsilon_\Lambda>0$ are both arbitrarily small. 

It is interesting to note that in the limit of \(\tilde{c}\to 0\), Eqs.~(\ref{eq:pressureprime}), (\ref{eq:massprime}), and (\ref{eq:alphaprime}) become the standard TOV equations within GR. 

\section{Results and discussions}
\label{RAD}

In this section, we discuss the binding energy of SNM and PNM as well as the EoS of SNM, PNM, and NS matter  predicted by standard RMF model with BSP parameter set ~\cite{SB2012-a,SB2012-b,SB2012-c,SB2012-d,SB2012-e} by taking into account the GUP correction within KMM model. We have found that the binding energy of SNM at low densities is sensitive to the change of GUP free parameter.  We also study the impact of GUP through  $\tilde{c}$ parameter of SL model on mass-radius of NS. We  also investigate in this section the sensitivity of recent mass-radius constraints from NICER to constraint GUP free parameter. For brevity, we do not show the units of either $\beta$ or $\tilde{c}$ in the figures. There we implicitly states that the units of $\beta$ and $\tilde{c}$ is, respectively, MeV$^{-2}$ and m$^2$. One of the relations discussed in Refs. \cite{Alonso-Serrano:2021} is $\Delta = 4\pi {\alpha_Q}^2$, where $\alpha_Q$ is related to $\beta$ by \(\beta = {\alpha_Q}^2 (l_P/\hbar)^2\) and $\Delta$ is related to the modified entropy formula for black hole from causal diamonds method 
\begin{equation}
	S_\text{mod}=(k_B A/4l_P^2)-(27\Delta/28)k_B \log(A/A_\text{min}).
\end{equation}
The causal diamond method is discussed in Ref.~\cite{Alonso-Serrano:2020dcz} as the basis to obtain the field equation of Serrano-Liska gravity theory, whose entropy formula is given by 
\begin{equation}
	S_\text{SL}=(k_B A/4l_P^2)+k_B\mathcal{C} \log(A/A_\text{min}), 
\end{equation}	
where $\mathcal{C}$ is related to $\tilde{c}$ by $\tilde{c}=\mathcal{C}l_P^2/18\pi$. Therefore, from comparing both entropy formula above, we have $\mathcal{C}=-(27/28)\Delta$, which implies the connection between $\beta$ and $\tilde{c}$ as 
	\begin{equation}
	\tilde{c} = -(3/14)\hbar^2\beta.\label{eq:parameterrelation}
	\end{equation}
It means negative value of $\tilde{c}$ microcanonical corrections \cite{Alonso-Serrano:2020dcz,Medved,Bhaduri:2003kv,Medved:2004gf,Chatterjee:2003su,Chatterjee:2004ji}. Note that the positive value obtained from canonical corrections due to thermal fluctuation \cite{Alonso-Serrano:2020dcz,Medved,Medved:2004gf,Chatterjee:2003su,Chatterjee:2004ji}. Therefore, from phenomenological view the sign of $\tilde{c}$ is determined by combining both corrections.

\begin{figure}
	\includegraphics[width=1.1\linewidth]{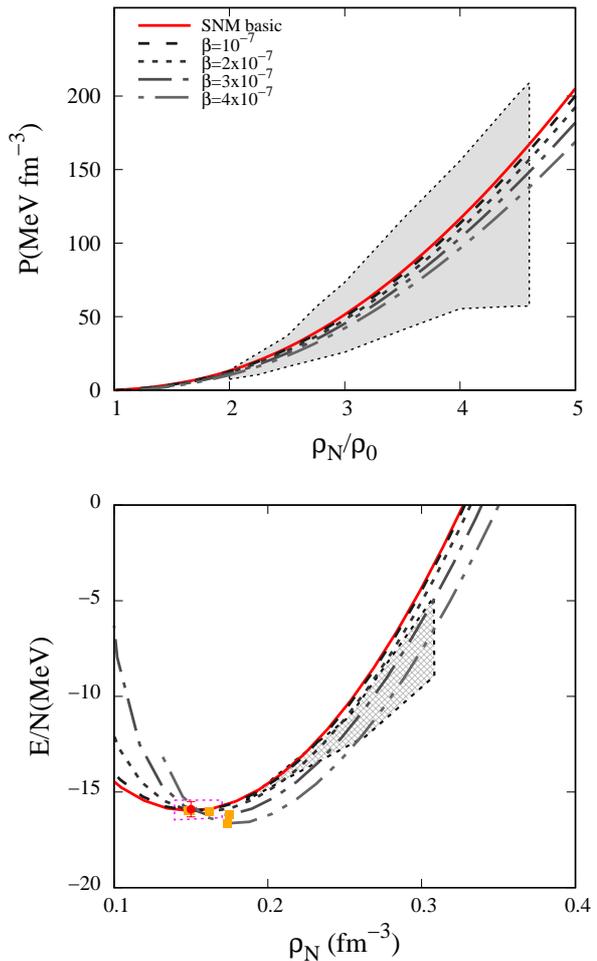}
	\caption{(Top) Our calculation results for EoS match the data from FOPI. But the results for the binding energy calculation are not trivial. (Bottom) Binding energy as a function of nucleon density, the grey shaded area is experimental data from FOPI \cite{LeFevre:2015paj}, the dashed box is the allowed binding energy at nuclear saturation density from Bethe-Weizacker mass formula. We also show the result from \cite{Brown:2013pwa} as the red dot with the error bar. The minimum value of the binding energy are also shown with box-shaped dots to show whose value is still inside the dashed box. \cite{Danielewicz:2002pu}. }
	\label{fig:plotSNM}
\end{figure}

\begin{figure}
	\includegraphics[width=1.1\linewidth]{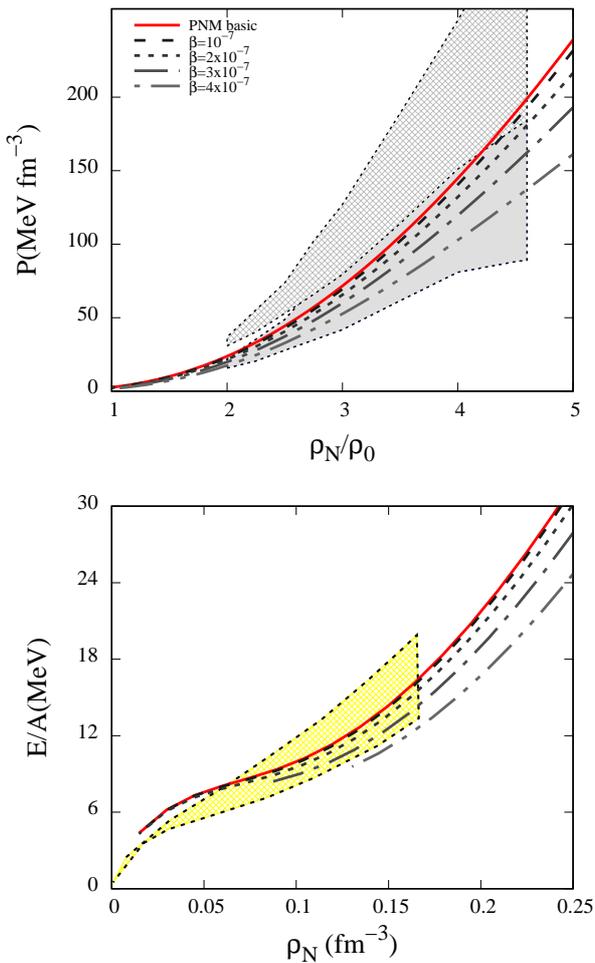}
	\caption{Similar to Fig.\ref{fig:plotSNM}, the results of varying $\beta$ in the case of pure neutron matter (PNM). (Top) the dark and light grey shaded area is a heavy-ion experiment from \cite{Danielewicz:2002pu}. (Bottom) The yellow shaded area is the theoretical binding energy from chiral effective field theory \cite{Kruger:2013kua}\cite{Holt:2016pjb}. }
	\label{fig:plotPNM}
\end{figure}

In Figs.~\ref{fig:plotSNM} and~\ref{fig:plotPNM}, we show the impact of $\beta$ variation on the  SNM and PNM  binding energies and EoSs. However, we have found that only for $\beta < 2.8\times 10^{-7}$ MeV$^{-2}$, our calculation does converge up to very low densities.  The reason is if we include too large value of  $\beta$ value ( $\beta \geq 2.8\times 10^{-7}$ MeV$^{-2}$) we can obtain the self-consistent solution of all meson equations only for relative moderate densities. Furthermore, for $\beta \geq 5\times 10^{-7}$ MeV$^{-2}$, the calculation can be done only for the nuclear matter density larger than the saturation density. Note that the shaded regions in the figures are extracted from experimental data. The EOS of PNM and SNM results,  are relatively in agreement with these data, except for the binding energy in the case of SNM, shown in the lower panel of Fig.~\ref{fig:plotSNM} is quite sensitive to $\beta$  value variation, specially in low density regions (densities less than saturation density of SNM). For  $\beta > 3.0\times 10^{-7}$ MeV$^{-2}$, the SNM saturation density is already out of box of the SNM binding energy constraint from Bethe-Weizacker mass formula.  Even not too significant, the $\beta$ has also effect in relative high densities regions. Increasing $\beta$ value leads to relative softer EoS in SNM and PNS as shown in top panels of Figs.~\ref{fig:plotSNM} and~\ref{fig:plotPNM}. 

In Table~\ref{tab:nucmatprop}, we have shown the impact of $\beta$ on SNM properties at saturation density. It can be observed that the $K_0$ and $K_{\rm sym}$ are sensitive to $\beta$ values. Even for $K_{\rm sym}$ change sign for $\beta > 2.0\times 10^{-7}$ MeV$^{-2}$ and $K_0$ value is already out side the constraint from heavy ion collision data \cite{Khan:2012} i.e. $230 \pm 40$ MeV. Therefore, we conclude that  nuclear matter properties could  constrain the $\beta$ value i.e.,  $\beta \le 2.0\times 10^{-7}$ MeV$^{-2}$.

\begin{table*}
	\centering
	\caption {Nuclear-matter properties at the saturation density $\rho_0$ predicted by BSP with various values for $\beta$. The Fermi momentum $k_F$, binding energy $E/N$, incompressibility coefficient for SNM $K_0$, symmetry energy $J$, $L$ and $K_{\rm sym}$ are defined in Eqs. (\ref{Eq:NMISCR}) and (\ref{Eq:NMIVEC}).}
	\label{tab:nucmatprop}
	\begin{tabular}{c | c  c c c c c}
		\hline\hline ~$\beta$~&~$10^{-7}$~  &~$2 \times 10^{-7}$~  &~$3 \times 10^{-7}$~&~$4 \times 10^{-7}$&~Constraints~  &~Refs~  \\\hline
		$E/N$ (MeV)                &-15.98  &-16.11  &-16.23 & -16.66  &-15.9 $\pm$ 0.4   &~\cite{Brown:2013pwa}  \\
		$K_0$ (MeV)               &254.70  &379.35  &428.99  & 530.81 &230 $\pm$ 40 &~\cite{Khan:2012} \\
		$J$ (MeV)                & 29.01 &30.12  &30.93 & 32.23 &31.7 $\pm$ 3.2 &~\cite{Li:2017,Oertel:2017}\\
		$L$ (MeV)               & 52.65 &59.64  &63.30 & 69.59 &58.7 $\pm$ 28.1 &~\cite{Li:2017,Oertel:2017}\\
		$K_{\rm sym}$ (MeV)               &6.29  &-4.89  & -11.09 & -27.43 & -400 $\le$ $K_{\rm sym}$ $\le$ 100  &~\cite{Cai:2017,Tews:2017,Zhang:2017}\\\hline\hline
	\end{tabular}\\
\end{table*}

\begin{figure}
	\includegraphics[width=1.1\linewidth]{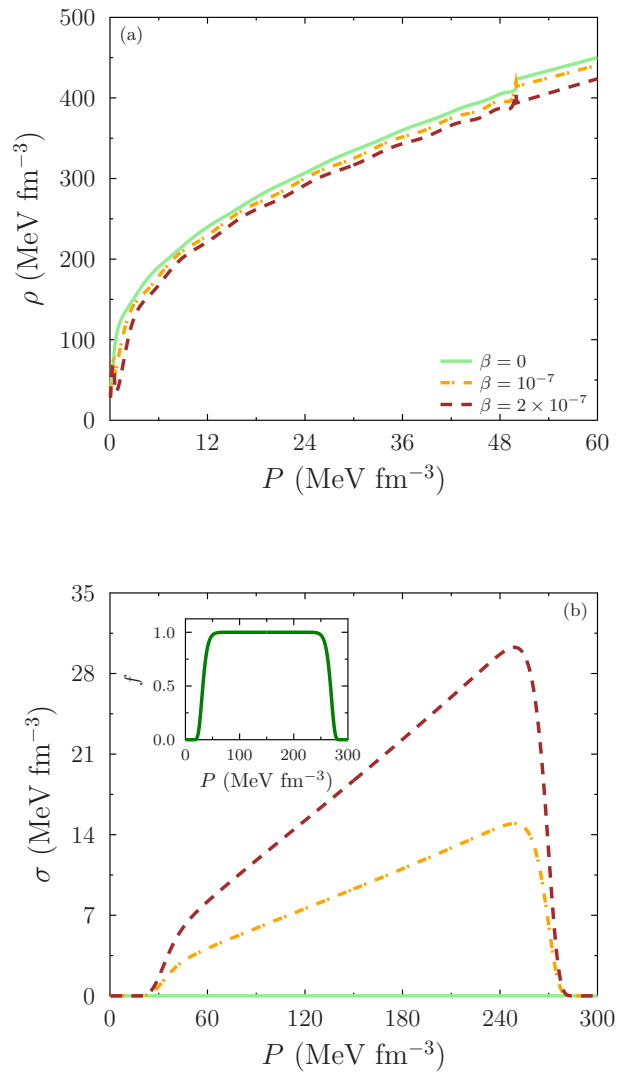}
	\caption{The EoS of NS shown in panel (a) and the anisotropy factor $\sigma$, shown in panel (b), is adjusted by multiplying $\sigma$ with $f=\exp[-(P/P_c-0.5)^{20}/(0.4)^{20}]$ (shown in the inset).}
	\label{fig:EoSsigmaGUP}
\end{figure}

In Fig.~\ref{fig:EoSsigmaGUP}, we show the EoS of NS matter and the anisotropy factor $\sigma$ of NS pressure. In the upper panel, the NS EoS consists of the crust EoS from Miyatsu {\it et al.}~\cite{Miyatsu:2013hea} and the core EoS is calculated using the standard RMF model. The analytical expression of all quantities to calculate NS core EoS is discussed in quite a detail in Sec.~\ref{RMF}. We have found that $\sigma$ profiles are increasing function, and its value is relatively large at high densities. This behavior could lead to instability in the center ($r \rightarrow 0$) of the star for the cases with a $\beta$ value relative large~\cite{Herrera:1997,Mak:2003,Setiawan:2019}. Therefore, to satisfy stability condition in NS center, $\sigma(P_c)\sim 0=\sigma_c$, we should adjust the anisotropy by $\sigma\to f\sigma$ by introducing phenomenologically cut-off function $f=\exp[-(P/P_c-0.5)^{20}/(0.4)^{20}]$ to force the $\sigma_c=0$ in the center, as is shown in panel (b) of Fig.~\ref{fig:EoSsigmaGUP}. Note that if we use the original anisotropy without any cut-off function, we could not have the numerical solution. It is evident that for $\beta=2\times 10^{-7}$ MeV$^{-2}$, the anisotropy  factor becomes significantly increased ($\sigma\sim 0.1P_c$). Furthermore, it is evident from the panel (a) of Fig. ~\ref{fig:EoSsigmaGUP} that the discrepancy in the EoS is not significant after introducing this cut-off. The reasons we do not have a TOV solution for  $\beta\gtrsim 10^{-6}$ MeV$^{-2}$  (large $\beta$) are twofold. First, the mesons equations in EoS are not convergent at low densities, and the $\sigma$ is too large in the regions near the center. This infers a constraint $\beta\leq 2\times10^{-7}$ MeV$^{-2}$ or ${\alpha_Q}^2\leq 10^{37}$ could be taken as the $\beta$ constraint from NS. Note that  ${\alpha_Q}^2=(\hbar/l_P)^2\beta$. To this end, it is worthy to note that the constraint obtained in this work is in agreement with a constraint from $^{87}$Rb cold-atom-recoil experiment~\cite{Giardino:2020myz}.

\begin{figure}
	\includegraphics[width=1.1\linewidth]{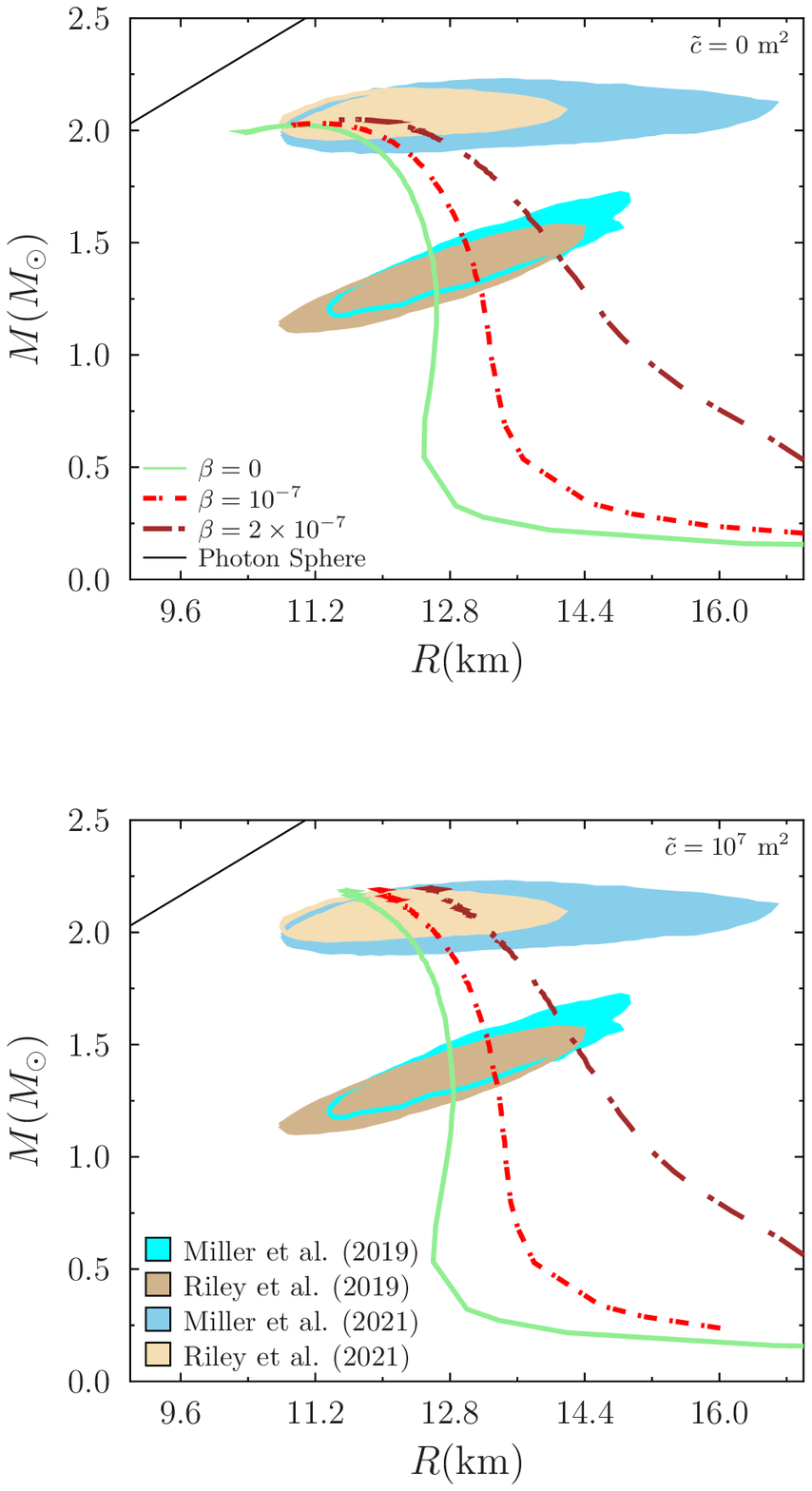}
	\caption{The mass-radius relation when $\beta$ value is varied. The data shape, which are NICER data of PSR J0030+0451 ($M\sim 1.4M_\odot$)~\cite{Miller:2019cac,Riley:2019yda} and PSR J0740+6620 ($M\sim 2.1M_\odot$)~\cite{Miller:2021qha,Riley:2021pdl}, are taken from Ref.~\cite{Li:2022okx}.}
	\label{fig:MRvarBeta}
\end{figure}

\begin{figure}
	\includegraphics[width=1.1\linewidth]{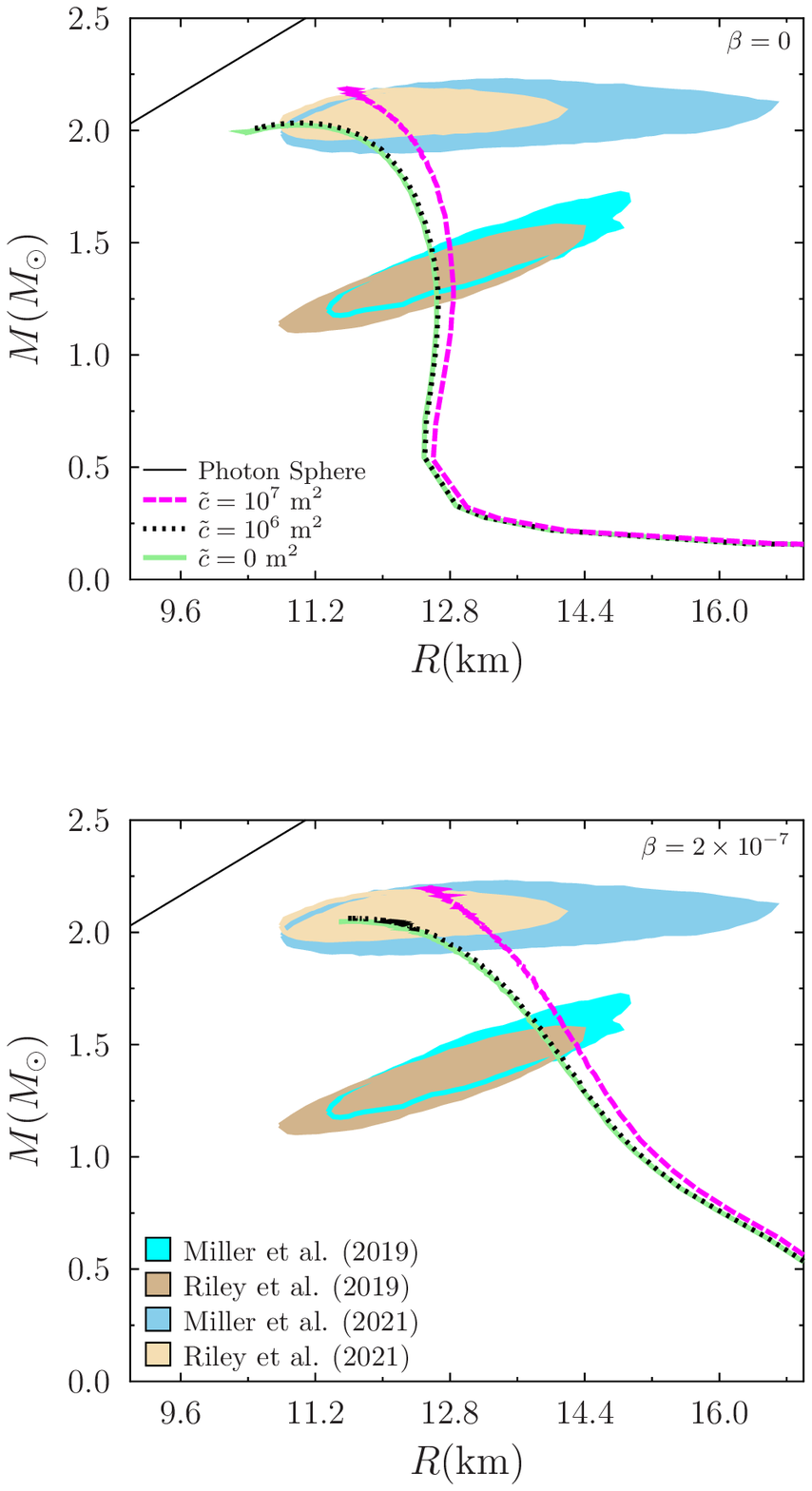}
	\caption{The mass-radius relation when $\tilde{c}$ value is varied. The data shape, which are NICER data of PSR J0030+0451 ($M\sim 1.4M_\odot$)~\cite{Miller:2019cac,Riley:2019yda} and PSR J0740+6620 ($M\sim 2.1M_\odot$)~\cite{Miller:2021qha,Riley:2021pdl}, are taken from Ref.~\cite{Li:2022okx}.}
	\label{fig:MRvarCt}
\end{figure}

In Figs.~\ref{fig:MRvarBeta} and~\ref{fig:MRvarCt}, we show the results of applying the EoS into the SL model. The former and the latter shows the variation of $\beta$ and $\tilde{c}$, respectively. Increasing $\beta$ value will decrease both the radius and mass, especially in the ``tail'' region when $P_c$ smaller than the value to obtain maximum mass. Increasing $\tilde{c}$ value will increase the maximum mass. In our calculation, we found that the numerical results behaves badly when either $\beta \gtrsim 10^{-6}$ MeV$^{-2}$ or $\tilde{c} > 10^7$ m$^2$. The reason for former case is due to $|\sigma| \sim P_c$ and for the latter is due to $|\tilde{P}|\gg |P|$.

Thus in this paper we have found a constraint for each parameter, i.e. $\beta \leq 2\times10^{-7}$ MeV$^{-2}$ and $\tilde{c} \leq 10^7$ m$^2$, if we consider bith parameters are independent. The former can be readjusted to be ${\alpha_Q}^2=(\hbar/l_P)^2\beta \leq 10^{37}$, which is is still in agreement to the $^{87}$Rb cold-atom-recoil experiment constraint. However latter gives a very large $\tilde{c}$ value, which conflicts with the fact that $\tilde{c}$ should be came from additional logarithmic term in the BH entropy.

It should be mentioned that $\tilde{c}$ and $\beta$ are related \cite{Alonso-Serrano:2021}. The relation can be seen from the modified entropy from both GUP and minimal area modification (see Eqs. (7) and (34) in Ref. \cite{Alonso-Serrano:2021}).
Since our results that yields a reasonably small anisotropy $\sigma$ should satisfy $\beta \leq 2\times 10^{-7}$ MeV$^{-2}$, we obtain $\tilde{c} > -16\times 10^{-34}$ m$^2$ from \eqref{eq:parameterrelation}. From our numerical results, we predict that using negative $\tilde{c}$ will lower the maximum mass in the mass-radius curve. However, according to our results which we use $\tilde{c}$ positive, $\tilde{c} < 10^7$ m$^2$ yields no significant shift in the mass-radius curve compared to the TOV GR result. Therefore, GUP modification on the matter leads to a much larger shift than SL modification on the geometry. This result is unsurprising because of two reasons: (1) $\tilde{c}$ should be in the order of $l_P^2$ and (2) if we assume that the effect of the SL model is as significant as the effect of the GUP model, then Refs. \cite{Giardino:2020myz,Okcu:2021oke,Das:2021lrb,Scardigli:2019pme} have shown that the upper bound for the GUP coupling constant value tends to be higher if the estimation is discussed in the case of gravity.

\section{Conclusions}
\label{sec_conclu}

We have shown that the upper bound for the GUP parameter, which modifies the nuclear matter properties and the NS matter, is $\beta = 2\times10^{-8}$ MeV$^{-2}$ and the upper bound for the SL parameter, which modifies the Einstein field equation, is $\tilde{c} = 10^7$ m$^2$. By employing $\beta =2\times10^{-7}$ MeV$^{-2}$ and $\tilde{c} = 10^7$ m$^2$, we obtain the mass-radius relation that satisfies NICER data for both PSR J0740+6620 ($M\sim 2.1M_\odot$) and PSR J0030+0451 ($M\sim 1.4M_\odot$). Our GUP parameter upper bound agrees with the $^{87}$Rb cold-atom-recoil experiment constraint. If we identify the entropy from both GUP and SL model and use our GUP parameter upper bound, we obtain that the resulting SL parameter lower bound is $\tilde{c} > -16\times 10^{-34}$ m$^2$. This lower bound is $10^{-40}$ smaller than the SL parameter upper bound by considering both parameters are independent, which is unsurprising because the upper bound for the GUP parameter tends to increase as we go from the quantum regime to the gravity regime.

%%%%%%%%%%%%%%%%%%%%%%%%%%%%%%%%%%%%%%%%%%%%%%%%%%%%%%%%%%%%%%%%%%%%%%%%
\begin{acknowledgements} 
IHB acknowledges the Sampoerna University internal grant 011/IRG/SU/AY.2021-2022.
\end{acknowledgements}



\end{document}